\documentclass[preprint]{aastex}
\usepackage{graphicx}
\usepackage{bm}
\usepackage{epsfig}
\usepackage{graphicx}
\usepackage{subfigure}
\usepackage{amsmath}
\usepackage{natbib}
\usepackage{multirow}

\shorttitle{Super Strong NS Magnetic Field in BeXBs}

\shortauthors{C. Shi et al.}

\begin{document}

%\linenumbers

\title{Super Strong Magnetic Fields of Neutron Stars in Be X-Ray Binaries Estimated with New Torque and Magnetosphere Models}

%\begin{CJK*}{UTF8}{gbsn}
\author{Chang-Sheng Shi\altaffilmark{1,2,3}, Shuang-Nan Zhang\altaffilmark{1,4 \star}, Xiang-Dong Li\altaffilmark{3,5}}

\altaffiltext{1}{National Astronomical Observatories, Chinese Academy of Sciences, Beijing 100012, China}
\altaffiltext{2}{College of Material Science and Chemical Engineering,
Hainan University, Hainan 570228, China}
\altaffiltext{3}{Key Laboratory of
Modern Astronomy and Astrophysics (Nanjing University), Ministry of
Education, Nanjing 210093, China}
\altaffiltext{4}{Key Laboratory of Particle Astrophysics, Institute of High Energy Physics,
Chinese Academy of Sciences, Beijing 100049, China; zhangsn@ihep.ac.cn}
\altaffiltext{5}{Department of Astronomy, Nanjing University, Nanjing 210093, China}
%\end{CJK*}
\email{$\star$ zhangsn@ihep.ac.cn}

\begin{abstract}
We re-estimate the surface magnetic fields of neutron stars (NSs) in Be X-ray binaries (BeXBs)
with different models of torque, improved beyond Klus et al. (2014). In particular, a new torque
model (Dai \& Li 2006) is applied to three models of magnetosphere radius. Unlike the
previous models, the new torque model does not lead to divergent results for any fastness
parameter. The inferred surface magnetic fields of these NSs for the two
compressed magnetosphere models are much higher than that for the uncompressed magnetosphere
model. The new torque model using the compressed-magnetosphere
radius (Shi, Zhang \& Li 2014) leads to unique solutions near spin equilibrium in all
cases, unlike other models that usually give two branches of solutions.
Although our conclusions are still affected by the simplistic assumptions about the magnetosphere radius calculations, we show several groups of possible surface magnetic field values with our new models when the interaction between the magnetosphere and the infalling accretion plasma is considered.
The estimated
surface magnetic fields for NSs BeXBs in the Large Magellanic Cloud, the
Small Magellanic Cloud and the Milk Way are between the quantum critical field and the maximum ``virial" value by the spin equilibrium condition.
\end{abstract}


\section{Introduction}
Be X-ray binaries (BeXBs) comprise a neutron star (NS) and a fast rotating Be-type main sequence star and are the most numerous class of high-mass X-ray binaries (HMXBs; Liu et al. 2006). Generally, a $8 M_{\odot} - 18 M_{\odot}$ Be star in a BeXB is surrounded by a circumstellar disk (Knigge et al. 2011; Reig 2011), where $M_{\odot}$ is the mass of the sun. A circling NS accretes matter from the disk arising from the wind of the
Be star, when it passes through the circumstellar disk and so most BeXBs
are transient systems due to the eccentric orbits ($e > 0.3$) (Reig 2011).

The accretion process of an NS in a BeXB has been discussed by many researchers such as Shakura
et al. (2012) and Ho et al. (2014). For X-ray binaries (XBs), a necessary condition for the formation of an accretion disk is that the specific angular momentum of gravitationally captured matter is
sufficiently high, or the accretion wind will persist (Burnard et al. 1983; Shakura et al. 2012, 2014). Reig (2011) discussed that Be stars as fast rotators might eject photospheric
matter with sufficient energy and angular momentum due to some weak processes, such as gas pressure or
non-radial pulsations, so that a Keplerian disk is formed. Porter \& Rivinius (2003) considered that emission lines and infrared excess in BeXBs might originate from an equatorial disk. James (2010) listed 14 HMXBs with
quasi-periodic oscillations, and many researchers considered that QPOs in XBs
 were an indication of an inner accretion disk around an NS
 (Miller et al. 1998; Stella \& Vietri 1998; Osherovich \& Titarchuk 1999; Abramowicz
 \& Kluz¡äniak 2001; Zhang 2004; Li \& Zhang 2005; Shi \& Li 2009, 2010; Shi
2011; Shi, Zhang \& Li 2014, etc.).

Giacconi et al. (1973 ) first observed the pulse period's variation of Hercules X-1 and later two states (spin-up or spin-down) of NSs were observed (Nagase 1989;
Bildsten et al. 1997).  Many authors (such as Ghosh \& Lamb 1979; Wang 1995; Klu¡äzniak \& Rappaport 2007) proposed different physical mechanisms in order to explain the change of the spin period of the NS in an XB. The changing spin period ($P$) of the NS in an XB was attributed to the change of the angular momentum of the NS due to accretion or outflow.

In most views, the corotation radius is considered as the dividing line between the spin-up state and the spin-down state for NSs in XBs. In the early literature such as
Ghosh \& Lamb (1979), the accretion matter from the donor star rotates around a central NS utill it reaches the inner disk terminal radius ($r_{\rm i}$), where the last angular velocity is the same as the angular velocity of the NS. The accretion disk has always been considered to be truncated at the magnetosphere radius ($r_{\rm m}$),
where the pressure of the accreting material is compensated by the magnetic pressure of
the magnetosphere (Pringle \& Rees 1972). Ghosh \& Lamb (1979)
supposed that the magnetic field lines were threaded in the Keplerian accretion disk in a broad transition zone, which has been extended to derive the torque acting on the NS.

Illiaronov \&
Sunyaev (1975) proposed that accretion could be continued only when the
accreting pulsar's magnetosphere rotates more slowly than the Keplerian angular frequency ($\Omega_{\rm k} = \sqrt{GM/r^3}$) of the accretion matter at $r_{\rm m}$, where $G,\ M,\ r$ are the gravitational constant, the mass of an NS, the radius to the center of the NS, respectively. In other word, if the magnetospheric radius lies outside the corotation radius ($r_{{\rm co}} = \sqrt[3]{GMP^2 / 4\pi ^2}$), at which the Keplerian angular frequency of the accretion matter is equal to the NS spin angular frequency ($\Omega_{\rm s}$), the accretion would be inhibited, i.e., the propeller effect emerges in the system (Shvartsman 1970). The remaining state of the NS (neither spin-up nor spin-down) is called spin equilibrium and it was believed to emerge when $r_{{\rm co}} = r_{\rm m}$ (Pringle \& Rees 1972, Klus et al. 2014).

Klus et al. (2014) obtained the spin period, change rate of spin period ($\dot{P}$), and
X-ray luminosity for 42 BeXBs with 2 states (spin-up or spin-down)
in the Small Magellanic Cloud (SMC). An unexpected conclusion was obtained:
the magnetic fields of the NSs in BeXBs are either much stronger or lower than those in low mass X-ray binaries (LMXBs). Ho et al. (2014) continued to discuss the two different kinds of solutions and they concluded that the strong magnetic field solutions from the non-spin equilibrium condition are more compelling because these solutions are close to
the solutions from the spin equilibrium condition.
They believed that spin equilibrium of BeXBs finally united different neutron star populations.

This paper is organized as follows. In Section 2, we review five available torque models and present a new torque model. In Section 3, we estimate the surface magnetic fields of NSs in BeXBs with these torque models and by assuming spin equilibrium for disk accretion. In Section 4, we give a discussion on the surface magnetic field of NSs in BeXBs. In the last section, we summarize this research.

\section{Models on dimensionless torque}

\subsection{Five available models of dimensionless torque}
The changing spin period of an accretion X-ray pulsar was discovered for a long time ago and it is widely accepted that the accretion torque acting on a NS system leads to the change of the spin period, i.e., \begin{equation}
\label{eq1}  - \dot{P} = NP^2 / 2\pi I_{{\rm eff}}\ ,
\end{equation}
where $N$ is the total torque acting on an NS system and $I_{{\rm eff}}$ is the effective moment of inertia of the NS (Ghosh et al. 1977). Ghosh \& Lamb (1979)
first made a magnetically threaded disk model (hereafter GL) to explain the changing spin period of NSs in XBs. Wang (1995) obtained three kinds of dimensionless torques in different
conditions. Klu\'{z}niak \& Rappaport (2007) obtained the results on torques acting on the
central star and the disk luminosity produced by the released gravitational
potential energy and the mechanical energy input to the disk via the magnetic torques; those results are independent of the viscosity prescription.
We will review the main results of the models below.

\subsubsection{The Model of Ghosh and Lamb}
Ghosh \& Lamb (1979) considered a magnetic NS with the dipolar magnetic field accreting matter from a Keplerian disk.
The steady accretion flow has an axial symmetry configuration and the magnetic moment is aligned with the rotation axis of the NS. They found that the magnetic coupling between the NS and the plasma outside $r_{{\rm i}}$ was extremely important for the physical mechanism of the changing spin of the NS in accretion magnetic XB. The inner disk terminal radius ($r_{\rm i}$) in GL can be written
as follows
 \begin{equation}
\label{eq2}
r_{\rm i} \approx 0.52 r_{\rm A} = 0.52{(\frac{\mu ^4}{2GM{\dot{M}^2}})}^{1 / 7}\ ,
\end{equation}
where $r_{\rm A}$ is the Alfv\'en radius for spherical accretion, $\dot{M}$ the accretion rate, and $\mu$ the magnetic moment of an NS.

 They expressed the total torque $N = N_{{\rm 0}} + N_{{\rm mag}}$, where $N_{{\rm 0}}$ is the material
 torque acting on the NS system when all the angular momentum of the accretion matter is transferred to the NS and $N_{{\rm mag}}$ is the magnetic torque produced by the magnetic coupling. The two variables can be expressed as follows
\begin{equation}
\label{eq3} N_{\rm 0} = \dot{M}\sqrt {GMr_{\rm i}}\ ,
\end{equation}
and
\begin{equation}
\label{eq4} N_{\rm mag} = - \int_{r_{\rm i} }^{_{r_{{\rm out}} } } {r^2} B_{\rm z} B_\varphi dr\ ,
\end{equation}
where $r_{{\rm out}}$ is the outer radius of the transition zone in the model of Ghosh \& Lamb and it
is always substituted by infinity because
$r_{{\rm out}} \gg r_{\rm i}$. The toroidal magnetic field ($B_\varphi$) is supposed
to be generated from the vertical magnetic field ($B_{\rm z}$) due to the differential rotation between an NS and its circling disk. The vertical magnetic field is supposed as a dipole magnetic field (i.e. $B_{\rm z} = \mu / r^3$) in all the models except Wang's models below, where $\mu$ is the magnetic moment of an NS.

The total torque can also be expressed as follows
\begin{equation}
\label{eq5} N = n\ast N_{\rm 0}\ ,
\end{equation}
where $n$ is a dimensionless accretion torque and it can be expressed approximately by
\begin{equation}
\label{eq6} n = 1.39\{1 - \omega [4.03(1 - \omega )^{0.173} - 0.878]\}(1 - \omega )^{ -
1} \ ,
\end{equation}
where $\omega$ is the fastness parameter defined as
\begin{equation}
\label{eq7} \omega \equiv \frac{\Omega _{\rm s} }{\Omega _{\rm k} (r_{\rm i}) } = (\frac{r_{\rm i} }{r_{{\rm co}} })^{3 / 2}\ .
\end{equation}
Equation (6) is accurate to within $5{\%}$ for $0 < \omega <0.9$  and the fastness parameter can be in the range $0 < \omega < 1$ in GL if the inner radius could be extended all the way to the center of the NS, but the real inner radius must be more than the radius of an NS and this restriction is adopted in this study below.

There are several problems in the model. (1) Wang (1987, 1995) pointed out that in GL the $B_\varphi$ component was overestimated so that the magnetic pressure exceeds the gas pressure in the outer region of the disk.
 (2) The zero torque will emerge when $\omega \sim 0.349$ or $\omega \sim 0.987$, which means that the torque reverses twice between a positive torque
and a negative torque when the accretion rate increases from zero to
a certain value (corresponding to $\omega < 0.349$), i.e., two spin-up and spin-down cycles should be observed during a monotonic change of luminosity over a large range. However, such a phenomenon has not been observed so far.

\begin{center}
\begin{figure*}[htp]
\label{fig1}
 \includegraphics[width=0.95\columnwidth]{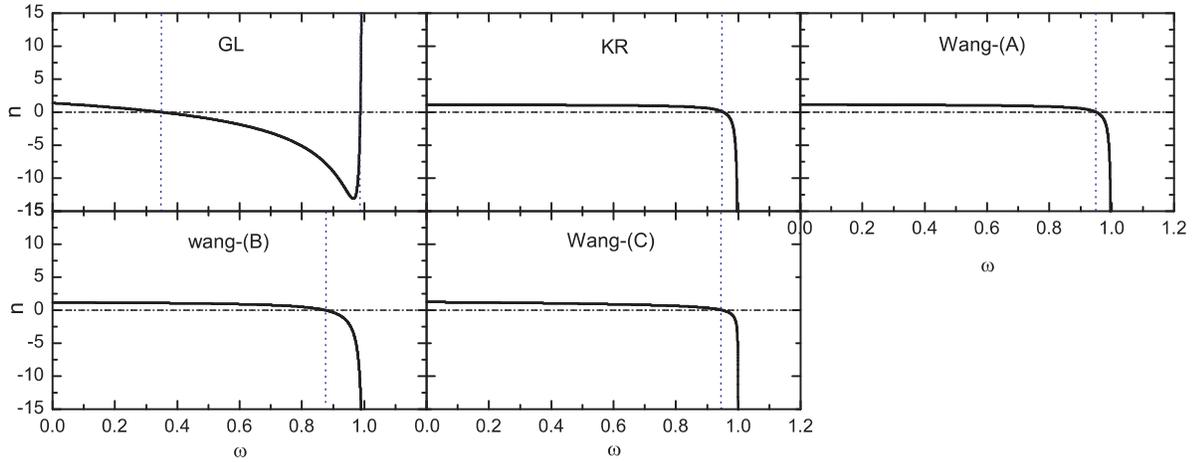}
 \caption{Relations between the dimensionless torque and the fastness parameters in five models. The dotted lines show the fastness parameters corresponding to zero torque. }
\end{figure*}
\end{center}

\subsubsection{Three models of Wang (1995)}
Wang (1995) gives a relation between the azimuthal magnetic field and the generated toroidal magnetic field due to the differential rotation as follows
 \begin{equation}
\label{eq8} \frac{B_\varphi }{B_z } = \gamma \tau _\varphi (\Omega _{\rm s} - \Omega _{\rm k} )\ ,
\end{equation}
where $\gamma$ is a numerical parameter and $\tau _\varphi$ is the dissipation timescale of $B_\varphi$. Then Wang (1995) discussed
three conditions that the dissipation timescale is determined by the reconnection outside the disk, the turbulent diffusion in the disk, and the Alfv\'en velocity, respectively. The three models in the three conditions are called Wang-(A), Wang-(B), Wang-(C) sequentially below.
The relations between $B_\varphi$ and $B_{\rm z}$ can be simplified when all the characteristic parameters are adopted (Wang 1987, 1996) as follows
\begin{equation}
\label{eq9}
 \left\{ {{\begin{array}{*{20}c}
 {{\rm (A)}:{B_\varphi}/{B_{\rm z}} = \Omega_{\rm s} / \Omega_{\rm k}- 1\ {\rm for}\ \Omega_{\rm s} < \Omega_{\rm k}; \ 1 - \Omega_{\rm k} / \Omega_{\rm s}\ {\rm for}\ \Omega_{\rm s} > \Omega_{\rm k}},\hfill \vspace{1mm}\\
 {{\rm (B)}:\ {B_\varphi}/{B_{\rm z}} = \Omega_{\rm s} / \Omega_{\rm k}- 1}\ , \hfill \vspace{1mm}\\
 {{\rm (C)}:\ |{B_\varphi}/{B_{\rm z}}| =  |\sqrt{{(1 - \Omega_{\rm s} / \Omega_{\rm k})\sqrt{4\pi p} B_z^{-1}}}|}\ , \hfill \\
\end{array} }} \right.
\end{equation}
where $p$ is the thermal pressure of the gas in a standard $\alpha$ disk (Shakura \& Sunyaev 1973).  The thermal pressure can be expressed as
\begin{equation}
\label{eq10}
p = 1.23 \ast 10^8 R_6^{4/5} M_{1.4}^{-5/2} L_{37}^{4/5} r_{6}^{-51/20} (1-1.12 r_{6}^{-1/2})^{4/5} {\rm gcm^{-1}s^{-2}}\ ,
\end{equation}
where $R$ denotes the radius of an NS, $L$ the X-ray luminosity; and the subscripts 6, 37 and $1.4\odot$
express the quantities in units of $10^6$ cm, $10^{37}\ {\rm erg s^{-1}}$, and
1.4 times the mass of the Sun, respectively. A negative vertical magnetic field (i.e. $B_{\rm z} = - \eta \mu / r^3$) is adopted to correct the direction of $B_{\varphi}$ in these models, where $\eta$ is a screening coeffeicient
as a constant ($\eta \leq 1$, Ghosh \& Lamb, 1979) and the characteristic value $\eta = 1$ is adopted in this study. The surface magnetic field of the NS in a BeXB would be amplified by a factor of about 2 when the extremely small value $\eta = 0.2$ was adopted.
 The inner boundary of the Keplerian disk in Wang's models (inside $r_{\rm co}$) can be obtained according to
\begin{equation}
\label{eq11}
({B_\varphi}{B_{\rm z}})|_{r = r_{\rm i}} = - \dot{M} \sqrt{GMr_{\rm i}}/(2r_{\rm i}^3)\ .
\end{equation}
The fastness parameter is restricted as $0 < \omega < 1$ in the above Equations (9, 11) and we can
simplify Equation (9) at the inner radius of Keplerian disk as follows
\begin{equation}
\label{eq12}
 {\frac{B_\varphi}{B_{\rm z}}}{\mid_{r = r_{\rm i}}} =\left\{ {{\begin{array}{*{20}c}
 {{\rm (A)}:\ \omega -1 }\  ,\hfill \\
 {{\rm (B)}:\ \omega -1}\ , \hfill \\
 {{\rm (C)}:\ - \sqrt{{(1 - \omega)|B_z^{-1}|\sqrt{4\pi p}}}}\mid_{r = r_{\rm i}}\ . \hfill \\
\end{array} }} \right.
\end{equation}

The dimensionless torques can be written as follows
\begin{equation}
\label{eq13}
n = \left\{ {{\begin{array}{*{20}c}
 {{\rm (A)}:\  \frac{(7 / 6) - (4 / 3)\omega + (1 / 9)\omega ^2}{1 - \omega }}\ ,\hfill \vspace{3mm}\\
 {{\rm (B)}:\  \frac{(7 / 6) - (4 / 3)\omega }{1 - \omega }}\ , \hfill \vspace{3mm}\\
 {{\rm (C)}:\  1 + \frac{1}{3}\frac{\omega ^{57 / 40}}{(1 - \omega )^{1 / 2}}[\int_\omega
^1 {\frac{(1 - y)^{1 / 2}}{y^{97 / 40}}dy - } \int_1^\infty {\frac{(y -
1)^{1 / 2}}{y^{97 / 40}}dy} ]}\ . \hfill \\
\end{array} }} \right.
\end{equation}

\subsubsection{The Model of Klu\'{z}niak and Rappaport}
Klu\'zniak \& Rappaport (2007) also discussed the properties of an accretion disk surrounding a magnetized
star and they first incorporated another component of the material torque ($-\dot{M}\Omega_s R^2$) omitted in other models, where $R$ is the radius of an NS (hereafter we will mark the model with KR). That component can be considered as a torque from the remaining angular momentum of the accreted plasma after the plasma falls on the surface of the NS and corotates with it.

This model predicted a dimensionless torque for $B_\varphi = B_z (1-\Omega_k/\Omega_s)$ inside $r_{\rm co}$ as follows
 \begin{equation}
\label{eq14}
n = \frac{10}{9} - \frac{1}{18}\frac{\omega }{1 - \omega }.
\end{equation}
The relation between $\omega$  and $r_{\rm A}$ can be obtained as follows
 \begin{equation}
\label{eq15}
\frac{1}{2\sqrt{2}} = (r_{\rm A}/r_{\rm co})^{7/2}\omega^{-10/3}(1-\omega)\ ,
\end{equation}
where the fastness parameter must be less than unity.

As shown in Figure 1, the relation between the dimensionless torque and the fastness parameter is very similar to Wang (A) and Wang (B). KR also has a very similar fastness parameter for the zero torque with Wang (C).

All of the above models imply that the fastness parameter is less than unity and it means that the accretion process is not in a propeller state. However, in reality the fastness parameter can also be slightly higher than unity because the accretion in the propeller phases has been observed (Finger et
al. 1994). As shown in Figure 1, the dimensionless torques in the above models are divergent when $\omega \rightarrow 1$ (also see Equation (29) in Wang 1987), i.e., the models result in an infinite torque.
The reason is that, e.g., Wang (1987) assumed that within the boundary layer of the accretion disk the toroidal field changes with the same law as that on the accretion disk, which is not verified.

\subsection{A New torque model}

Besides the torque equilibrium condition (i.e. the method of balancing the magnetic and viscous torques), we can also estimate the inner disk terminal radius by the pressure (or energy) equilibrium condition. Both theoretical and numerical investigations (e.g., Lynden-Bell \& Boily 1994; Lovelace et al. 1995; Uzdensky et al. 2002) have shown that when the toroidal field becomes comparable to the poloidal field, the fields open up and the star-disk linkage is broken. Therefore, in the boundary layer the magnetic pitch is always around unity. Inserting this condition into Equation (28) in Wang (1987), one can derive the expression of the magnetospheric radius, which is the same as that in the case of pressure/energy balance. This is exactly what Wang did in his later paper (Wang 1996) to estimate $r_{\rm m}$ by torque balance and the divergence problem can be avoided naturally. Numerical calculations by Long et al. (2005) also confirmed that the magnetospheric radii in both cases coincide with each other under this condition. Dai \& Li (2006) thus adopted the inner disk terminal radius by the energy equilibrium condition and made a new torque model. In this subsection, we extend this new model to the inner disk terminal radius considered as three kinds of magnetosphere radii from the pressure (or energy) equilibrium condition.

\subsubsection{Basic prescriptions}
We adopt the typical assumptions of NSs in BeXBs as follows.
(1) The spin of an NS with a dipolar magnetic field is aligned to the magnetic axis and perpendicular to the Keplerian accretion disk. (2) The azimuthal component of the magnetic field is generated by rotation shear when the poloidal component keeps the dipolar field unchanged. (3) The magnetosphere is force-free and reconnection takes place outside the disk (Wang 1995). (4) The interaction between the NS and the accretion disk mainly comes from the magnetic field anchored on the NS surface. (5) The accretion process is steady. (6) The accreted matter at the magnetosphere radius corotates with the NS and the magnetosphere, i.e, the accretion matter keeps the last angular momentum at the magnetosphere radius.
Here the magnetosphere radius ($r_{\rm m}$) is considered to be the inner radius of the Keplerian dsik.
The total material torque acting on the NS system can be described as
\begin{equation}
\label{eq16}
N_{{\rm mat}} = \dot{M}\sqrt{GM{r_{\rm m}}} - \dot{M}\Omega_{\rm s} r_{\rm m}^2 = \dot{M}\sqrt{GM{r_{\rm m}}}(1 - \omega)\ ,
\end{equation}
and so the total torque can be expressed as follows
\begin{equation}
\label{eq17}
N = N_{{\rm mat}} + N_{\rm mag}\ .
\end{equation}
Wang (1995) found
\begin{equation}
\label{eq18}
\frac{B_\varphi }{B_z } = \left\{ {{\begin{array}{*{20}c}
 {\gamma (\Omega _{\rm  s} / \Omega _{\rm  k} - 1)},\ \Omega _{\rm  s} \leq \Omega _{\rm  k}\ , \hfill \\
 {\gamma (1 - \Omega _{\rm  k} / \Omega _{\rm  s} )},\ \Omega _{\rm  s} > \Omega _{\rm  k}\ , \hfill \\
\end{array} }} \right.
\end{equation}
where $\gamma$ is a constant and hereafter we suppose $\gamma=1$, which means the highest steepness of the transition between Keplerian motion inside the disk and the corotation with the NS outside the disk (Wang  1995; Dai \& Li 2006). The magnetic torque in Equation (4) can be integrated by combining Equation (17) and we obtain
\begin{equation}
\label{eq19}
N_{\rm  mag} = \left\{ {{\begin{array}{*{20}c}
 {\frac{1}{3}\mu ^2r_{\rm  m}^{ - 3} (\frac{2}{3} - 2\omega
 + \omega ^2),\ \omega \le 1}\ ,\hfill \\
 {\frac{1}{3}\mu ^2r_{\rm  m}^{ - 3} (\frac{2}{3}\omega ^{ - 1} -
1),\ \omega > 1}\ ,\hfill \\
\end{array} }} \right.
\end{equation}
where $\mu = B R^3$ is the magnetic moment and the original expression in Dai \& Li (2006) in the parentheses of the above first sub-equation was $1 - 2\omega + \frac{2}{3}\omega ^2$ due to typo.

Combing Equations (5), (16), (17), and (19), we obtain
\begin{equation}
\label{eq20}
n = \left\{ {{\begin{array}{*{20}c}
 {(1 - \omega ) + \frac{1}{3}\frac{\mu ^2r_{\rm  m}^{ - 3} }{\dot{M}
 \sqrt {GMr_{\rm  m} } }(\frac{2}{3} - 2\omega + \omega ^2),\ \omega \le 1}\  ,\hfill \\                                                                                                                                                                                                                                                                  {(1 - \omega ) + \frac{1}{3}\frac{\mu ^2r_{\rm  m}^{ - 3}}
 {\dot{M} \sqrt {GMr_{\rm  m}} }(\frac{2}{3}\omega ^{ - 1} - 1),\ \omega > 1}\ .\hfill \\
\end{array} }} \right.
\end{equation}

\subsubsection{Three Models of Magnetosphere Radius}
Lamb et al. (1973) discussed the spherical accretion
process onto a compact star and gave a definition of the magnetosphere radius.
Many authors also explored the outer boundary of the magnetospheres of pulsars (e.g., Elsner \& Lamb 1977; Burnard et al. 1983; Mitra 1992;
Li \& Wang 1995; Weng \& Zhang 2011) for the disk accretion process
and they obtained analogous conclusions according to the spherical
accretion. The general expression of that radius in those models can be written
as follows
 \begin{equation}
\label{eq21}
r_{\rm m} = k r_{
\rm A}\ .
\end{equation}
Klus et al. (2014) adopted Alfv\'en radius, which can be expressed as follows
 \begin{equation}
\label{eq22}
r_{\rm m1} \equiv r_{\rm A} = 3.69\ast 10^8R_6 ^{10 / 7}M_{1.4\odot} ^{1 / 7}B_{12} ^{4 / 7}L_{37} ^{ - 2
/ 7}\ {\rm cm}\ ,
\end{equation}
where $B$ denotes the surface magnetic field of an NS; and the subscripts 12 express the quantities in units of $10^{12}\ \rm G$ .

Shi et al. (2014) extended the definition of the magnetosphere radius and they considered that the magnetohydradynamic equation, including gravity, barometric pressure,
magnetic pressure, and inertial force acting on the accretion plasma at the magnetosphere radius in a corotation frame of reference, must be satisfied. They considered that the magnetic field is compressed by
the accretion matter utill the accretion matter reaches the magnetosphere
radius but the dipolar magnetic field inside the magnetosphere
is not compressed.
They obtained a magnetosphere radius where the part of the magnetic field of an NS outside $r_{\rm  m}$ is compressed according to the conservation of the magnetic flux at the accretion disk plane. The expression of the magnetosphere radius is
 \begin{equation}
\label{eq23}
r_{\rm m2} = 5.54\ast 10^7 \alpha _{0.1}^{4 / 15} R_6^{206 / 135}
M_{1.4 \odot }^{ - 1 / 135} B_{12}^{16 / 27} L_{37}^{ - 34 / 135}
f^{ - 136 / 135}\ {\rm cm} \ ,
\end{equation}
where $\alpha$ is the viscosity parameter
and $\alpha_{0.1} =  \alpha/0.1$,
${f = (\mbox{1} - \sqrt {\frac{R}{r}} )^{1 / 4}}$.

Besides the above magnetosphere radius from theoretical modeling, Kulkarni \& Romanova (2013) found from numerical simulation
another dependence of the magnetosphere radius on the surface
magnetic field, the mass, the radius of an NS and the
luminosity of a XB.
In their three-dimensional magnetohydrodynamic simulation of
an accretion process at the quasi-equilibrium state, the gravitational,
centrifugal, and pressure gradient forces were considered. It is different from Shi, Zhang \& Li (2014) in that the globular magnetic field is
compressed by the accretion matter, i.e., the magnetosphere within the magnetosphere radius
is also compressed.
They obtained a magnetosphere radius as follows
\begin{equation}
\label{eq24}
r_{{\rm m3}} = 7.12 \ast 10^7 R_6^{13 / {10}} M_{1.4 \odot }^{1 / {10}}
\ast L_{37}^{ - 1 / 5} B_{12}^{2 / 5}\ {\rm cm}.
\end{equation}
Please refer to Figure 2 in Shi, Zhang \& Li (2014) for comparisons of
the three models of magnetosphere radius.

\subsubsection{The New Torque Model}
As seen from Section 2.2.1, the expression (Equation (20)) of the dimensionless torque will not be changed if we adopt the different expressions of the magnetosphere radius, but only a true magnetosphere radius can lead to a correct dimensionless torque. The three kinds of magnetosphere radii in Subsection 2.2.2 are all admitted.
In order to compare the magnetic field of an NS in BeXBs with the uncompressed magnetosphere to the one with the compressed magnetosphere below, we substitute the three expressions of magnetosphere radius (see Equations (22)-(24)) into Equation (20). Then we obtain
\begin{equation}
\label{eq25}
{\rm (1) for}\ \omega \le 1,\  n = \left\{ {{\begin{array}{*{20}c}
 r_{\rm m1}:\  (1 - \omega ) + \frac{\sqrt 2}{3}(\frac{2}{3} -
2\omega  + \omega^2 )\ ,\hfill \\
 r_{\rm m2}:\  (1 - \omega ) + 314.258 \ast f^{34 / 10} P_1^{ - 1 / 12} L_{37}^{ - 3 / {20}}
 \omega^{ - 1/{12}} ({\frac{2}{3}} - 2\omega  + \omega^2 )\ , \hfill \\
 r_{\rm m3}:\  (1 - \omega ) + 543.248 \ast P_1 \omega ({\frac{2}{3}} - 2\omega  + \omega^2 )\ , \hfill \\
\end{array} }} \right.
\end{equation}

\begin{equation}
\label{eq26}
{\rm (2) for}\ \omega > 1,\  n = \left\{ {{\begin{array}{*{20}c}
 r_{\rm m1}:\  (1 - \omega ) + \frac{\sqrt 2}{3}(\frac{2}{3}\omega ^{ - 1} -
1)\ ,\hfill \\
 r_{\rm m2}:\  (1 - \omega ) + 314.258 \ast f^{34 / 10} P_1^{ - 1 / 12} L_{37}^{ - 3 / {20}}
 \omega^{ - 1/{12}} (\frac{2}{3}\omega ^{ - 1} -
1)\ , \hfill \\
 r_{\rm m3}:\  (1 - \omega ) + 543.248 \ast P_1 \omega (\frac{2}{3}\omega ^{ - 1} -
1)\ . \hfill \\
\end{array}}} \right.
\end{equation}

As we can see in Figure 2, unlike the five-dimensionless torque models
in subsection 2.1, our result is convergent when the magnetosphere radius is close to the corotation radius (also see Equations (28) and (29)). Like some other dimensionless torque models, the zero torque acting on an NS is also not at the spin equilibrium state due to the assumption that the interaction between the magnetic field and the accretion disk also exists when $r_{\rm  i} = r_{\rm  m} = r_{\rm  co}$, i.e., $\omega = 1$ (see Equations (3), (23), and (24)). It should be noted that the dimensionless torque for $r_{\rm  m2}$ and $r_{\rm  m3}$ do not depend only on the fastness parameter ($\omega$), but are also related to $P$
(for $r_{\rm  m2}$ and $r_{\rm  m3}$) and $L$ (for $r_{\rm  m2}$), respectively.

The dimensionless torques for the magnetosphere radii of the compressed
magnetic fields ($r_{\rm  m2}\ \&\ r_{\rm  m3}$) change
much more rapidly than the dimensionless torque for $r_{\rm  m1}$ and
the reason can be found from Equation (27).
The surface magnetic fields of NSs for $r_{\rm  m2}\ \&\ r_{\rm  m3}$
change with magnetosphere radius more rapidly than that for $r_{\rm  m1}$
(see Figure 2 in Shi, Zhang \& Li 2014).
The values of the negative dimensionless torque from Equation (27)
change only with $\mu^2$ when we compare the dimensionless torque in Figure 2 with the same fastness parameter.
So the rapidly changed surface magnetic fields can lead to the large
difference of the three panels of Figure 2.
The last two panels of Figure 2 in fact express a relation between
the surface magnetic field of NSs and the fastness parameters.

\begin{center}
\begin{figure*}[htp]
\label{fig2}
 \includegraphics[width=1.0\columnwidth]{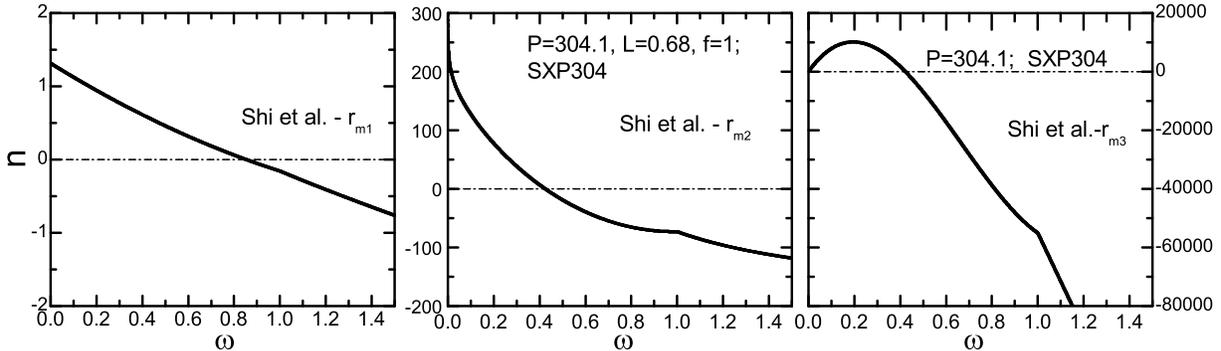}
 \caption{Relation between the dimensionless torque and the fastness parameter for
$r_{{\rm m1}}$, $r_{{\rm m2}}$, and $r_{{\rm m3}}$, respectively.}
\end{figure*}
\end{center}

\section{Estimating the Surface Magnetic Fields of NSs in BeXBs}
Ghosh \& Lamb (1979) first calculated the magnetic moment of several XBs in the range $3\ast10^{29}  - 4\ast10^{32}\ {\rm G}\cdot{\rm cm}^3$ by their dimensionless torque model (GL). Klus et al. (2014) computed the
magnetic fields of NSs in BeXBs in SMC by GL and KR with the Alfv\'en radius (i.e. $r_{\rm m1}$) and they obtained two kinds of solutions (the upper or lower solutions).
Ho et al. (2014) argued that the upper solutions reflect the real magnetic field of NSs in BeXBs.

Here we  we extend the
calculation on the surface magnetic fields of the NSs in BeXBs for
Wang's three-dimensionless torque models (Wang 1995) and the model (KR) in disk accretion.
In order to compare the characteristics of the sources in disk accretion with that of the sources whose accretion modes are unknown, we calculate the the magnetic fields for all sources from Klus et al. (2014) except SXP 701 ($\dot{P} = 0$ for SXP 701) with a disk accretion condition.
Klus et al. (2014) computed the results for GL and KR with $r_{\rm A}$,
but the inner radius was obtained as Equation (15), not $r_{\rm A}$ in KR.
Here we recalculate the surface magnetic field of NSs in BeXBs with the inner radius obtained by Equation (15).
We adopt all the sources from Klus et al. (2014) except the source (SXP 701) in subsections 3.1 and 3.2 and also the BeXBs in the Milk Way (MW),
the Large Magellanic Cloud (LMC), and SMC in subsection 3.3 because
we do not know the spin period derivatives of many sources in LMC
and MW.

\subsection{Estimating Surface Magnetic Field With Four Available Torque Models}
In this work we select the characteristic values of NSs in accretion X-ray pulsars, i.e., $M = 1.4M_\odot$, $R = 10^6\ {\rm cm}$, $\alpha = 0.1$ (King et al. 2007) and so the inertial moment of a spherical NS is $1.12 \ast 10^{45}\ \rm{g \cdot cm^2}$.  From Equations (1), (3), and (5), we obtain
\begin{equation}
\label{eq27}
n\omega ^{1 / 3}\sqrt {r_{\rm co} } = - 2.72\ast 10^8\ast
\dot{P} P^{ - 2}L_{37} ^{ - 1}I_{45} M_{1.4\odot}^{1 / 2} R_6^{ - 1}\ ,
\end{equation}
where the unit of $\dot{P}$ is $\rm s\ yr^{-1}$. Then the fastness parameter ($\omega$)
can be obtained when the dimensionless torque (Equation (13) or (14)) is substituted into Equation (27).

We then compute the magnetic field solutions of NSs in BeXBs when Equations (7) and (10)-(12) are solved using $B_{\rm z} =  \mu / r^3$, and the surface magnetic fields of NSs in BeXBs in Wang's models can be described as follows
\begin{equation}
\label{eq28}
B = \left\{ {{\begin{array}{*{20}c}
 {{\rm (A)}:\  6.05 \ast 10^{10}r_8^{7 / 4} M_{1.4\odot} ^{ - 1 / 4}L_{37}^{1 / 2} R_6^{- 5 / 2} {(1 - \omega )^{- 1 / 2}}}\ {\rm G},\ {\rm for}\ \omega < 1\ ,\hfill \vspace{3mm}\\
  {\  \  \ 6.05 \ast 10^{10}r_8^{7 / 4} M_{1.4\odot} ^{ - 1 / 4}L_{37}^{1 / 2} R_6^{- 5 / 2} {(1 - \omega^{-1} )^{- 1 / 2}}}\ {\rm G},\ {\rm for}\ \omega > 1\ ,\hfill \vspace{3mm}\\
 {{\rm (B)}:\   6.05 \ast 10^{10}r_8^{7 / 4} M_{1.4\odot} ^{ - 1 / 4}L_{37}^{1 / 2} R_6^{- 5 / 2} {(1 - \omega )^{- 1 / 2}}}\ {\rm G}\ ,\hfill \vspace{3mm}\\
 {{\rm (C)}:\  1.56\ast 10^{12}R_6 ^{- 7 / 3}M_{1.4\odot} ^{- 1 / 3}\ast L_{37} ^{2 / 3}\ast
r_{\rm co, 8} ^{4 / 3}\omega ^{8 / 9}(1 - \omega )^{ - 1 / 3}p_{1}^{ - 1 / 6}}\ {\rm G}\ , \hfill\\
\end{array} }} \right.
\end{equation}
where $r_{\rm co, 8} = r_{\rm co} / (10^8\ \rm{cm}$), $p_{1} = p / (1\ {\rm dyn} /{\rm cm}^2$).

Similar to the above processes, we can obtain the surface magnetic field of NSs in BeXBs in KR with Equations (7), (14), (15), (22), and (27)
\begin{equation}
\label{eq29}
B = 1.50\ast 10^{11}L_{37}^{1 / 2} M_{1.4\odot}^{1 / 3} R_6^{ - 5 / 2} P_1^{7
/ 6}\omega ^{5 / 3}\ast(1 - \omega )^{ - 1 / 2}\ {\rm G}\ ,
\end{equation}
where $P_1 = P/1\ {\rm s}$.

\begin{center}
\begin{figure*}[h]
\label{fig3}
 \includegraphics[width=0.9\columnwidth]{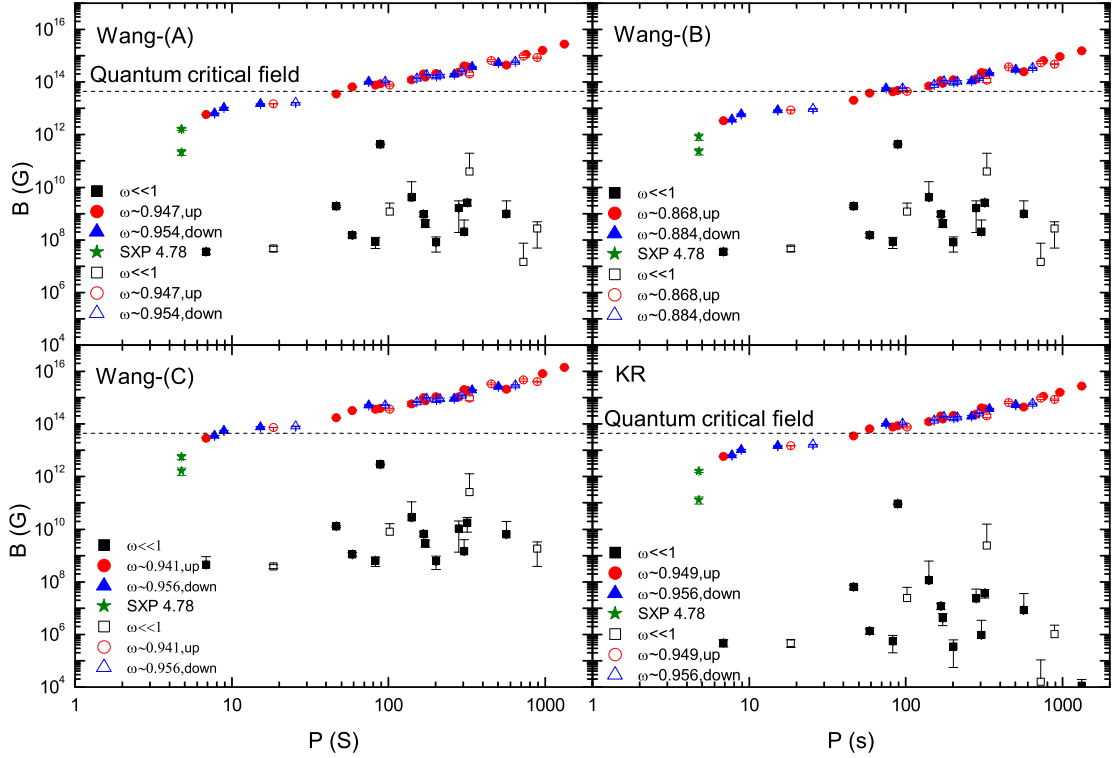}
 \caption{General results of Equations (28) and (29) between $B$ and $P$ for four-dimensionless torque models (Wang A, B, and C and KR). Hereafter in all figures
  $R^\ast$ indicates the radius of an NS, ``up" indicates ``spin-up" and ``down" indicates ``spin-down"; the dashed lines mark
  the quantum critical field, i.e., $B = 4.414  \ast 10^{13}\ {\rm G}$;
  the data for SXP 4.78 with a star. The filled symbols express the sources in disk accretion and the open symbols express the sources in unknown accretion modes, as identified in Klus et al. (2014).
 }
\end{figure*}
\end{center}

As shown in Figure 3, we show the results of the NSs in BeXBs
for Wang's models and KR with their respective radius according to the different characteristic values of the fastness parameters. The detailed values of the fastness parameters for different sources are different, but the
upper branch of the solutions in one model corresponds to a similar fastness parameter and we mark the
mean values of the parameters in Figure 3. The fastness parameters for the lower branch of the solutions are very different and we only mark their scope ($\omega \ll 1$; the same process with Figure 4).
Most solutions with the fastness parameter ($\omega \sim 0.9$) are higher than the critical magnetic field, indicating the same property of all BeXBs. Besides those solutions, there are several special solutions in all the panels for the source (SXP 4.78) with different fastness parameters and there is no
magntic field solution for in GL (see Klus et al. 2014). There is no solution for the magnetic field for the sources (SXP 2.37, SXP 11.5, and SXP 16.6) in the above models.

\subsection{Estimating Surface Magnetic Field Using Our New Torque Model}
In this subsection we apply our new model presented in Section 2.2.3 and estimate the magnetic fields of NSs in BeXBs; note that here the magnetosphere radius outside
the radius of an NS is considered as the inner radius of the accretion disk.

Combing Equations (7) and (22)-(24), we can obtained the expressions of the fastness parameters. After combing Equation (25)-(27) with the expressions of the fastness parameters, we can obtain the surface magnetic fields of NSs in BeXBs as follows
\begin{equation}
\label{eq30}
B_{12} = \left\{ {{\begin{array}{*{20}c}
 r_{\rm m1}:\  5.04 \ast 10^{14} R_6 M_{1.4 \odot }^5 [( - \dot{P} / n)^{ - 1 / 2} P_1 L_{37} ^{ 3 / 7}]^{ - 7}\ ,\hfill \vspace{3mm}\\
 r_{\rm m2}:\ 3.70 \ast 10^{15} \alpha _{0.1}^{ - 9 / 20} f^{17 / 10} R_6^{4 / 5} M_{1.4
\odot }^{203 / 40} [( - \dot{P} / n)^{ - 1 / 2} P_1 L_{37}^{59 / 135}]^{ - 27 / 4}\ , \hfill \vspace{3mm}\\
 r_{\rm m3}:\  6.18 \ast 10^{22} R_6^{{25}/4} M_{1.4}^{11 / 4} [( - \dot{P} / n)^{ - 1 / 2}P_1 L_{37}^{9 / 20}]^{-10}\ . \hfill \\
\end{array}}} \right.
\end{equation}
As shown in Figure 4, our result shows the solutions of Equations (30) when the dimensionless torques (Equations (25) and (26)) are substituted. Similar to the results
of the above five-dimensionless torque models, the stronger magnetic field solutions
in our model also have a similar fastness parameter. It means that the magnetic field
has a relatively steady relation with the spin period when omitting
the small change of the luminosity in BeXBs, suggesting that these solutions are near spin equilibrium. The fastness parameters for these stronger field solutions are lower than that obtained with
the five-dimensionless torque models in subsection 2.1, which means that the equilibrium
radii are smaller here.
The lower solutions with different and very small fastness parameters
do not have a steady relation with the spin period, because these solutions are far from spin equilibrium.

About the two branches of solutions in the left (for $r_{\rm m1}$) and right panels (for $r_{\rm m3}$), the upper solutions are obtained for all these BeXBs, but the lower solutions are obtained only for some of them. There are two possibilities for this: (1) the upper and lower solutions give the correct magnetic fields for different BeXBs. However, in this case it is difficult to understand why the XBs have two very distinct groups of magnetic fields and there are almost no BeXBs with magnetic fields in between. (2) The lower solutions are non-physical and thus should be abandoned, and thus only the upper solutions are reasonable, as argued by Ho et al. (2014). In addition, all the lower solutions are for the spin-up NSs and have very small fastness parameters, i.e., far from
the spin equilibrium. It was observed that some spin-up NSs in XBs
sometimes also transit spin-down without significant luminosity change, implying they are not far from spin equilibrium.

In fact, in the middle panel (for $r_{\rm m2}$) of Figure 4, only the upper solutions are obtained in near spin equilibrium, which indicates that the model of $r_{\rm m2}$ is more reasonable than either $r_{\rm m1}$ or $r_{\rm m3}$. We therefore conclude that the upper solutions with the $r_{\rm m2}$ model give the correct magnetic fields of these NSs, which are also in near spin equilibrium.

Note that although our new model presented above is an improvement beyond Klus el at. (2014), our conclusions are still affected by the simplistic assumptions about the magnetosphere radius calculations. Better magnetosphere radius models are still needed for realistically estimating more the magnetic fields when the detailed interactions between the disk and the magnetosphere are fully considered and the compression effect of the magnetic field in the accretion process is better described.

\begin{center}
\begin{figure*}[htp]
\label{fig4}
 \includegraphics[width=1.0\columnwidth]{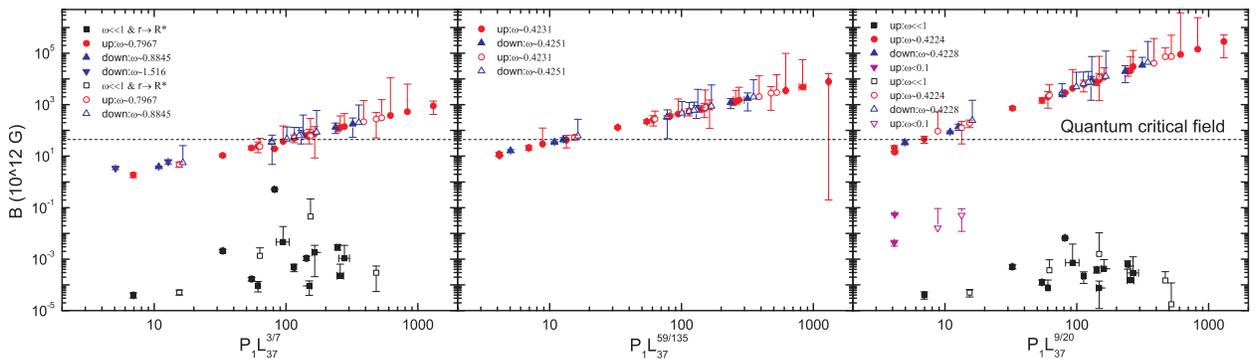}
 \caption{Relation between the surface magnetic field of the NS ($B_{12}$) and $PL^\delta$ for $r_{{\rm m1}}$, $r_{{\rm m2}}$, and $r_{{\rm m3}}$, respectively.  The filled symbols express the sources in disk accretion and the open symbols express the sources in unknown accretion modes, as identified in Klus et al. (2014).}
\end{figure*}
\end{center}

\subsection{Estimating the Surface Magnetic Field by Spin Equilibrium}
A spin-evolution timescale ($P/\dot{P}\sim 10^2 - 10^4\ \rm yr$) of XB pulsars in the SMC (SXPs) can be obtained from the data of Klus et al. (2014), which is much shorter than the $10^7\ {\rm  year}$ age
of an OBe star and the decay timescale $10^6 {\rm \ yr}$ of the global magnetic field in BeXBs (Ho et al. 2014).
Therefore, BeXBs with strong magnetic fields in the SMC have enough time to achieve near spin equilibrium and the surface magnetic fields of NSs in most BeXBs will not be
too low when they achieve the spin equilibrium.

The above models are all including non-zero derivatives of the spin period ($\dot{P} \neq 0$) and are called non-spin equilibrium models (Klus et al. 2014). We also can obtained the surface magnetic field
of an NS with the spin equilibrium condition ($r_{{\rm m}} = r_{{\rm co}}$) for $r_{{\rm m1}}$, $r_{{\rm m2}}$ and $r_{{\rm m3}}$ as follows
 \begin{equation}
\label{eq31}
B_{12} = \left\{ {{\begin{array}{*{20}c}
 {r_{{\rm m1}}:\ 0.25M_{1.4} ^{1 / 3}R_6 ^{ - 5 / 2}P^{7 / 6}L_{37} ^{1 / 2}}
\hfill \vspace{3mm}\\
 {r_{{\rm m2}}:\ 6.50\alpha _{0.1}^{ - 9 / 20} f^{17 /
10}M_{1.4 \odot }^{23 / 40} R_6^{ - 103 / 40} P^{9 /
8}L_{37}^{17 / 40} } \hfill \vspace{3mm}\\
 {r_{{\rm m3}}:\ 8.54M_{1.4 \odot } ^{7 / 12}R_6^{ - 13 / 4}
P^{5 / 3}L_{37}^{1 / 2} } \hfill \\
\end{array} }} \right.\ .
\end{equation}

In order to compare the solutions from spin equilibrium with those from non-spin equilibrium, we show all the solutions (the solutions from Section 3.1 by non-spin equilibrium models, the solutions from the new model in Section 3.2 for $r_{\rm m2}$ and the solutions from Equation 31 by spin equilibrium for $r_{\rm m2}$) in Figure 5. According to the criterion of disk accretion by Klus et al. (2014), we showed the solutions for the sources in disk accretion ($v_{\rm rel}/v_{\rm crel} \leq 1$ corresponding to the sources in disk accretion) in the left panel of Figure 5, and the solutions for the sources whose accretion mode has not been determined due to no available $v_{\rm rel}$ in the right panel, where $v_{\rm rel}$ is the relative velocity of accreted matter and $v_{\rm crel}$ is its maximum relative velocity for which disk accretion can take place.
We obtain the surface magnetic field of the NSs in the BeXBs in unknown accretion mode by the models of disk accretion. The result for the sources in wind accretion can be referred to Klus et al. (2014). As shown in Figure 5, the solutions with larger characteristic values for $\omega$ from the non-spin equilibrium condition are very close to the solutions from the spin equilibrium condition.

\begin{center}
\begin{figure*}[h]
\label{fig5}
 \includegraphics[width=1.0\columnwidth]{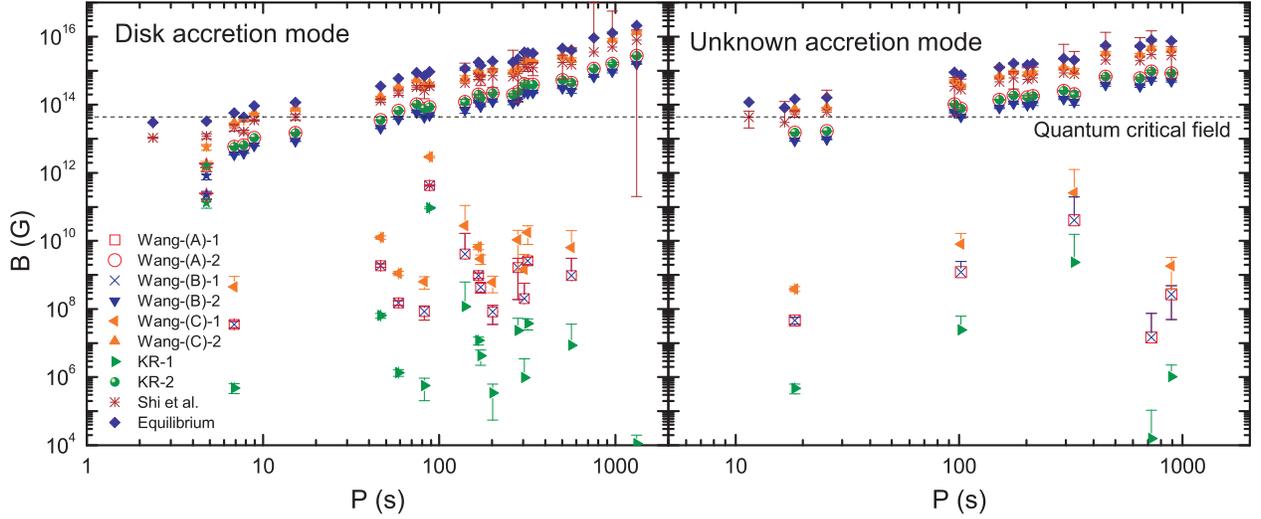}
 \caption{Composed figure of the surface magnetic fields of NSs from spin equilibrium and non-spin equilibrium conditions, in which the data come from Figure 3, our torque model for $r_{\rm  m2}$ and from the disk spin equilibrium condition for $r_{\rm  m2}$. Here ``Equilibrium" indicates the solutions from spin equilibrium,``1, 2" indicates the first and second solutions for some sources from non-spin equilibrium. The data marked with stars come from the special sources ( SXP 4.78). Left: the sources are in disk accretion. Right: the sources are in unknown accretion modes. }
\end{figure*}
\end{center}

Figure 6 shows the distribution of the spin frequency derivatives ($\dot{\nu}$) of NSs in BeXBs in the SMC, which shows that the values of most $\dot{\nu}$ are close to zero and thus most BeXBs are quite close to spin equilibrium. A lot of observations on the sway of the spin period of an NS between spin-up and spin-down (e.g. Chakrabarty et al. 1995; Bildsten et al. 1997; Nelson et al. 1997; Postnov et al. 2015) suggest that the accretion processes of NSs in BeXBs are very close to spin equilibrium. It is consistent with the conclusion from the above analysis.
\begin{center}
\begin{figure*}[h]
\label{fig6}
 \includegraphics[width=1.0\columnwidth]{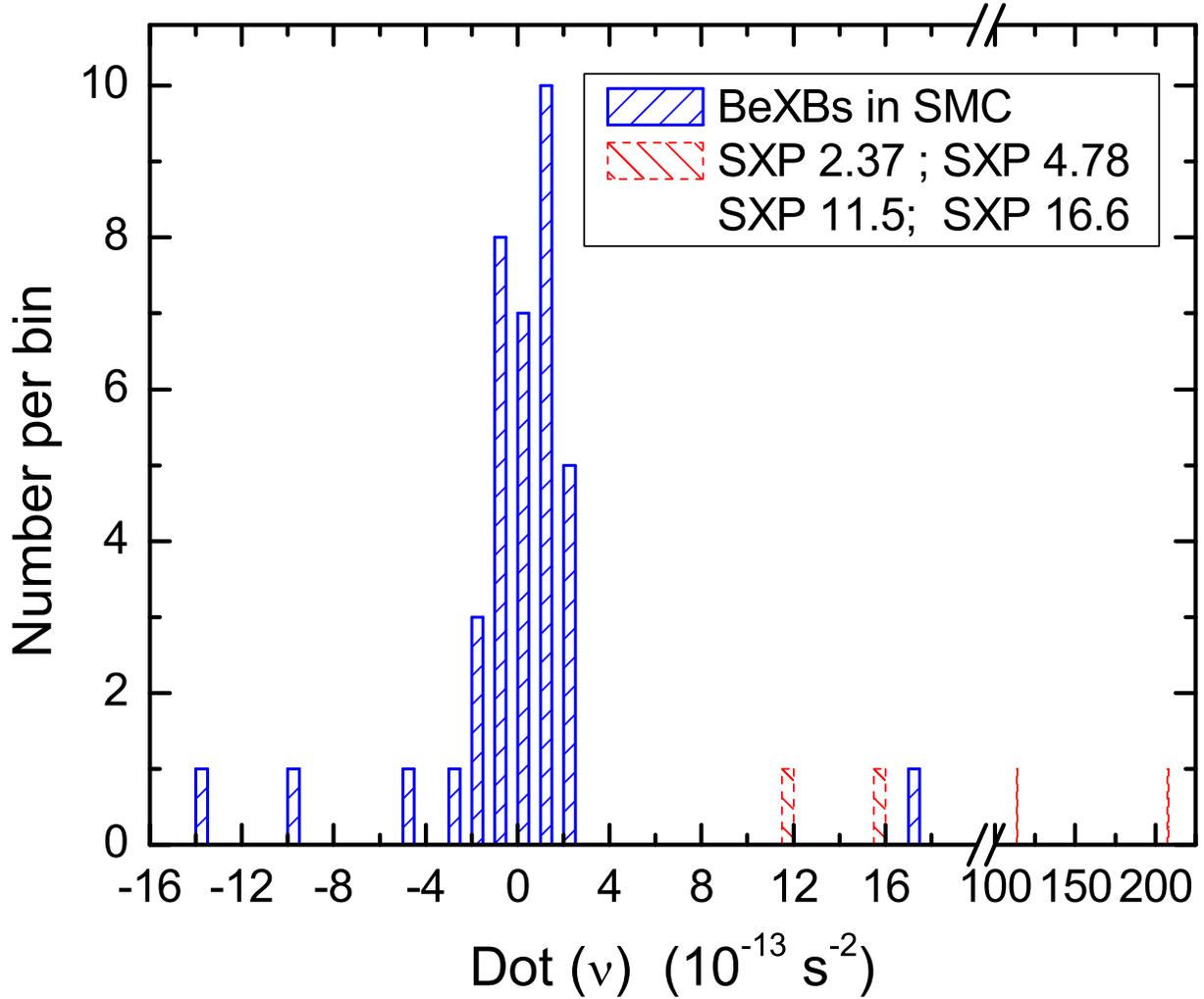}
 \caption{Distribution of the spin frequency derivative of the NSs in BeXBs in the SMC. }
\end{figure*}
\end{center}

Because most BeXBs are very close to spin equilibrium, we can compute the surface magnetic fields of the NSs in BeXBs in the MW, the LMC and the SMC in the spin equilibrium condition, as shown in Figure 7.
We also show the surface magnetic field calculated from the observed cyclotron lines
of some sources with $B_{12} = 0.863 \ast E_{10\ \rm kev}$, if we suppose that
the electrons are in the ground state and the line comes from the surface of an NS,  where $E_{10\  {\rm keV}}$ is the energy of the cyclotron absorption line in units of $10\ {\rm keV}$. The result suggests $10^{12} - 10^{13}\ \rm G$ surface magnetic fields of NSs in BeXBs, consistent to that reported by Caballero \& Wilms (2012).
If we consider the cyclotron line energy of protons in the same way, the surface magnetic field of an NS should be $B_{12} = 1586\ast E_{10\ {\rm keV}}$, which is much closer to the upper solutions discussed above. Maybe the discussed cyclotron lines are indeed produced by
protons. The result has also been discussed in the isolated magnetars by Harding \& Lai (2006).

The solutions of $r_{{\rm m2}}$ and $r_{{\rm m3}}$ have much higher magnetic fields than those of $r_{{\rm m1}}$ because of the compression effect of the magnetic field. If the compression effect of the magnetic field exists, the magnetic fields of most NSs in BeXBs will exceed the quantum critical magnetic field ($44.14\ {\rm TG}$), but all of them will not exceed the
maximum ``viral" value $1.77 \ast 10^{18} M_{1.4 \odot} R_6^{-2}\ \rm G$, which is defined according to the condition that the magnetic energy ($\frac{4\pi R^3}{3} \ast \frac{B^2}{8 \pi}$) equals the gravitational binding energy ($\frac{GM^2}{R}$). It seems
that there is a ceiling effect of the surface magnetic
fields of NSs in BeXBs due to the maximum ``virial" value as shown in Figure 7.

 \begin{center}
\begin{figure*}[htp]
\label{fig7}
 \includegraphics[width=1.0\columnwidth]{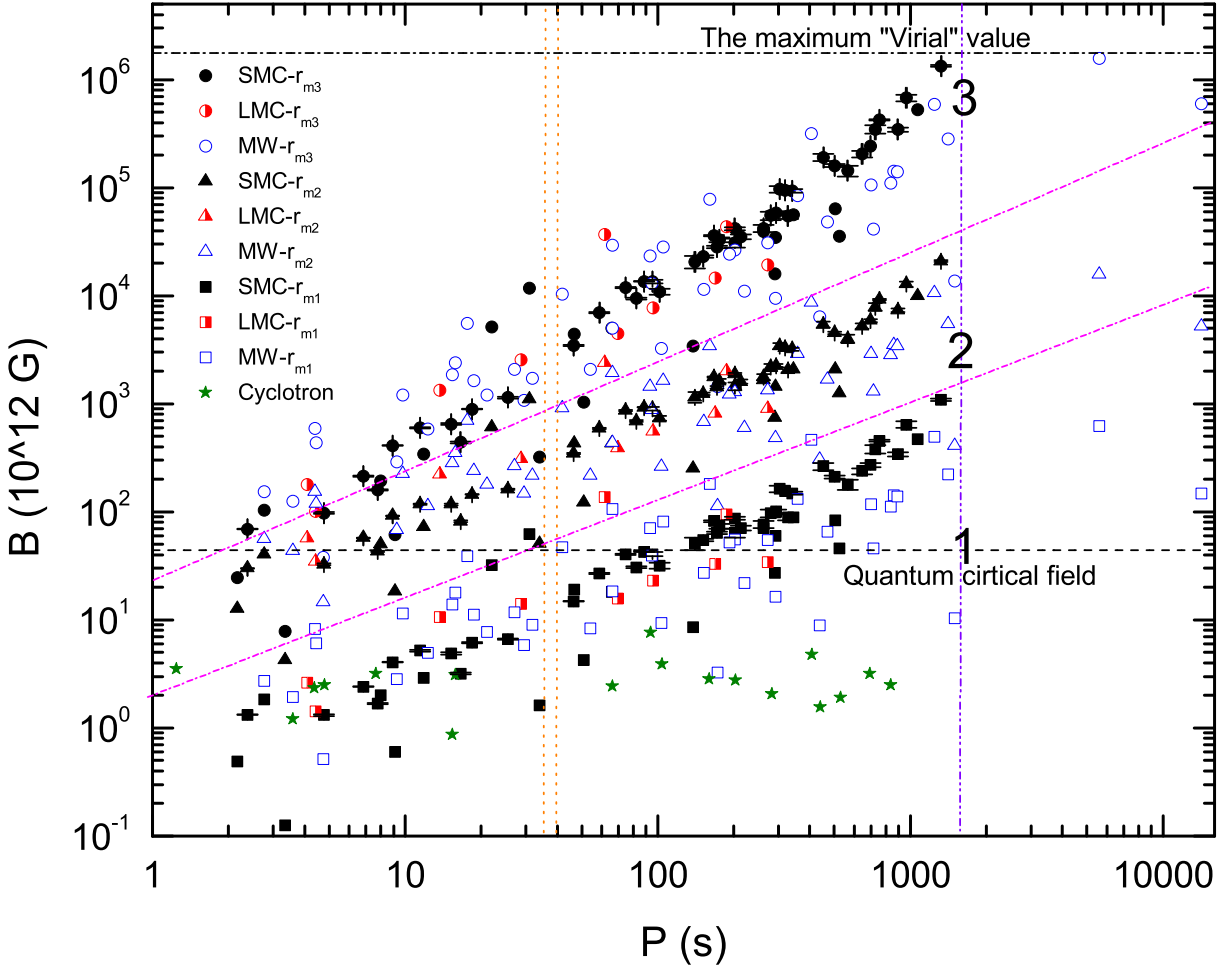}
 \caption{Relation between the surface magnetic fields of NSs in BeXBs in and the spin period of NSs. The stars mark the magnetic field inferred from the observed cyclotron
  line sources (data from http://www.sternwarte.uni-erlangen.de/wiki/doku.php?id=cyclo:start). The data on the spin period of NSs and luminosity are from Knigge et al. (2011), Reig (2012),
   Klus et al. (2014), Cheng et al. (2014), and references therein.
   The circle points indicate the solutions for $r_{{\rm m3}}$, the triangle points for $r_{{\rm m2}}$,
  the square points for $r_{{\rm m1}}$,
  the solid points indicate the solutions from the sources for the SMC, the half-filled points for the
  LMC, and the open points for the MW as well.}
\end{figure*}
\end{center}

\section{Discussion}
\subsection{Wind accretion}
Although many observations indicate that there might be an accretion disk in a BeXB (Porter \& Rivinius 2003, James 2010), a system containing an OBe star with
a non-truncated circumstellar disk might also be in the wind accretion state (Shakura
et al. 2012). Shakura et al. (2012) ascribed  the phenomenon of torque reversals to wind accretion; however, torque reversals have also been observed in disk-fed pulsars. Nevertheless, it is reasonable to assume that
an accretion X-ray pulsar swinging frequently at the zero torque should be in wind accretion state. Klus et al. (2014) found that the ratio $v_{\rm rel}/v_{\rm Crel}$ can be obtained for nearly half of the the systems in their sample of BeXBs and all of these systems are in disk accretion; the other half may contain some systems in wind accretion (also see Figure 5).

However all our calculations presented here are only valid for the disk-fed pulsars. The models of Wang (1995), Klu{\'z}niak \& Rappaport (2007) and Dai \& Li (2006) were all developed only for disk accretion. For pulsars not in spin equilibrium, our dimensionless torque model evolved from Dai \& Li (2006) and Wang (1995) cannot be extended to wind accretion; the main reason is that the model of magnetic field topology used by us is derived from disk accretion (Wang 1995). As a matter of fact, Klus et al. (2014) also did not calculate the surface magnetic field of the NSs of BeXBs not in spin equilibrium for wind accretion. Therefore, further development is still required to calculate the magnetic field of the NSs in BeXBs not in spin equilibrium for wind accretion.

For pulsars in spin equilibrium, we assumed that the corotation radius is equal to the magnetosphere radius for a disk-fed pulsar. However, this assumption is clearly in valid for wind accretion. Klus et al. (2014) assumed that the torques of spin-up and spin-down cancel each other for wind accretion and calculated their surface magnetic fields this way. The magnetic fields derived by Klus et al. (2014) from two spin equilibrium models on wind accretion are lower by approximately one or two orders of magnitude than that for disk accretion in spin equilibrium obtained by them or by us using the compressed magnetosphere, respectively.

Is it then possible that the NSs in BeXBs in wind accretion have systematically lower surface magnetic fields than those in disk accretion? We believe this is unlikely because the surface magnetic fields of NSs in BeXBs should be their
intrinsic properties at birth and thus not influenced significantly by their different accretion modes and processes at a relatively small age in comparison with the decay timescale of the magnetic field. Given the relative reliability and simplicity of disk accretion models, the super strong magnetic fields calculated with disk accretion models might reflect the common characteristic of all NSs in BeXBs if a significant fraction of the BeXBs are indeed in disk accretion. This is consistent with the fact that the identified disk-fed systems cover the same parameter ranges as other systems, as shown in Figures 3-5.

\subsection{Super Strong Magnetic Fields}

Most young NSs have magnetic fields with orders of magnitude $10^{12}\ \rm{G}$, but much lower magnetic fields with orders of magnitude $10^{8}\ \rm{G}$ are always reported in LMXBs (e.g. Shi, Zhang \& Li 2014). In contrast, much higher magnetic fields with orders of magnitude $10^{14}\ \rm{G}$ are reported in magnetars, including anomalous X-ray pulsars (AXPs) and soft gamma repeaters
(SGRs; Woods \& Thompson 2006, p. 547; Mereghetti 2008). The most effects of the magnetars are ascribed to the strong magnetic fields, and they are especially powered by the strong magnetic fields. If a young NS (with a strong magnetic field $10^{11}\sim10^{15}\ \rm G$ )
has an evolutionary connection with the LMXBs, the damping
timescale of the magnetic field is $10^9\ \rm year$ (Harding \& Lai 2006). Thus, it seems that
the lower magnetic field solutions with very small fastness parameters cannot be achieved
because the lower magnetic fields are much lower than the fields of LMXBs and the evolution
timescale should be much longer.
We found that the surface magnetic fields of most NSs in BeXBs are higher than
$10^{14}\ \rm G$.
Reig (2012) discussed the observation of the BeXB (4U 2206+54) and
they thought that BeXBs are accreting magnetars.
The NSs with strong magnetic fields in BeXBs as accreting magnetars
may supply a channel to evolve into isolated magnetars because the two kinds of NSs have very similar characteristics, such as
long spin periods and high spin period derivatives.

The inferred extremely strong magnetic fields of NSs in some BeXBs
supply a key clue
to studying some basic physical problems (see Harding \& Lai 2006).
(1) Photon splitting becomes dominant as an attenuation mechanism,
but photon splitting cascades may be prevented due to Adler's selection rules
(Harding \& Baring 1996; Harding et al. 1998). Thus, that
radiation from low energy photons in BeXBs should be absent or weak (e.g. radio radiation).
(2) The equation of state of the bulk interior of an NS may be affected by the magnetic field since the field
approaching the ``virial" value. The NSs in those BeXBs would have lower density
and the area of the radiation region would become larger
than the NSs with weak magnetic fields
because the high magnetic energy comparable with the gravitational potential
energy compensates the interaction among baryons.
The general relativity effect (such as gravitational redshift and the space-time curvature)
would be alleviated because of the smaller mass-radius ratio.
(3) Thomson \& Compton cross sections are strongly reduced for the photon energies
much lower than the cyclotron energy if the magnetic field is very strong
(Paczynski 1992), and the maximum luminosity of an accreting BeXBs may exceed the
Eddington luminosity. The ultraluminous X-ray sources (NuSTAR J095551+6940.8) as
a accreting magnetar in HMXBs (Bachetti et al. 2014; Ek{\c s}i et al. 2015)
may be an example of the BeXBs.

\section{Summary}
In this study we reviewed five-dimensionless torque models of X-ray binaries and presented a new model of the dimensionless torque based
on the calculation of Dai \& Li (2006). Using these models, we calculated the surface
magnetic fields of the NSs in BeXBs in SMC. Finally, we estimated the surface magnetic field according
to the spin equilibrium, including the NSs in BeXBs in the LMC and the Milky Way. Our conclusions can be summarized as follows.
\begin{enumerate}
\item{The dimensionless torque in the five previous dimensionless torque models is divergent when the magnetosphere radius is equal to the corotation radius.}
\item{The dimensionless torque in our new model is convergent and depends on the NS spin period and luminosity of X-ray binaries besides the fastness parameters, when applied to the two compressed magnetosphere radius models.}
\item{Two branches of solutions are found with all previous torque models, i.e., one with larger
fastness parameters very close to spin equilibrium
and the other with very small fastness parameters corresponding to spin-up far from spin equilibrium. The upper branch solutions with higher magnetic fields are close to spin equilibrium and more reasonable than that in the lower branch.}
\item{By applying our new dimensionless torque with the compressed magnetosphere radius model in Equation (23) (Shi, Zhang \& Li 2014), only one branch of solutions is obtained, corresponding to the upper solutions obtained in other models. This combination of dimensionless torque and magnetosphere models is thus favored.}
\item{By assuming spin equilibrium, the estimated surface magnetic fields of the NSs in BeXBs in the SMC, the LMC and the Milky Way show similar characteristics.}
\item{The estimated surface magnetic fields for the compressed magnetosphere models are much
higher than that for the uncompressed magnetosphere model. They exceed the
quantum critical magnetic field, but do not exceed the maximum ``virial" value.}
\end{enumerate}

%We calculated the surface magnetic fields of NSs in BeXBs by the spin equilibrium for the
%three magnetosphere radius models and those magnetic field solutions are very close to
%the solutions with the higher fastness parameters calculated by the non spin-equilibrium.

\acknowledgments

This work is supported by 973 Program of China
under grant 2014CB845802, by the National Natural Science Foundation
of China under grant Nos. 11133002, 11373036, 11203009, 11563003, 11133001, and 11333004, by the Strategic Priority Research Program ``The Emergence of Cosmological Structures" of the Chinese Academy of Sciences, Grant No. XDB09000000, by
the Qianren start-up grant 292012312D1117210, and by the 54th postdoctoral Science Foundation funding of China.

%\bibliographystyle{apj}
%\bibliography{APLib_chandra_v7_submit.bbl}
%\bibliography{ref_APLib_aasstyle.bib}
%\clearpage
%\bigskip
%\bigskip

\end{document}